\documentclass[showpacs,twocolumn,showkeys,amsmath,amssymb,pra]{revtex4} 

\usepackage{graphicx}   
\usepackage{dcolumn}    
\usepackage{bm}         
\begin{document} 
 
\title{Bose-Einstein condensates in a double well: mean-field chaos and multi-particle entanglement} 
 
\author{Christoph Weiss}
\email{weiss@theorie.physik.uni-oldenburg.de}
\author{Niklas Teichmann}
\affiliation{Institut f\"ur Physik, Carl von Ossietzky Universit\"at,
                D-26111 Oldenburg, Germany
}

\keywords{Chaos, mesoscopic entanglement, Bose-Einstein condensation, quantum Fisher information}
                  
\date{submitted: 06 October 2008, to be published in the proceedings of the Laser physics conference
in Trondheim 2008}
 
\begin{abstract}

A recent publication [Phys.\ Rev.\ Lett.\ 100, 140408 (2008)] shows that there is a relation between
mean-field chaos and multi-particle entanglement for BECs in a periodically shaken double well. ``Schr\"odinger-cat'' like mesoscopic superpositions in
phase-space occur for conditions for which the system displays mean-field chaos. 
In the present manuscript, more general highly-entangled states are investigated. Mean-field
chaos accelerates the emergence of multi-particle entanglement; the boundaries of stable
regions are particularly suited for entanglement generation.

\end{abstract} 
\pacs{03.75.Gg, 05.45.Mt, 74.50.+r, 03.65.Ta}

\maketitle


\newpage

\section{Introduction}
Systems of ultra-cold atoms with time-periodic potential differences have recently
started to be implemented experimentally both on the single-particle level~\cite{KierigEtAl08} and for
Bose-Einstein condensates (BECs)~\cite{LignierEtAl07,SiasEtAl08}. These experiments realise schemes of
tunnelling-control devised in recent years~\cite{GrossmannEtAl91,EckardtEtAl05}. Very recently, even the control of
the Mott-insulator transition via periodic shaking~\cite{EckardtEtAl05b,CreffieldMonteiro06}
has been realised experimentally~\cite{Zenesini08}. This would, roughly speaking, correspond to making
a glass of water freeze by turning on a loud-speaker. Proposals for shaking-induced tunnelling
control for ultra-cold atoms
not realised so far include schemes for entanglement
generation~\cite{TeichmannWeiss07,WeissTeichmann07,Creffield2007}. The tunnelling control
based on shaking the potentials uses concepts successfully implemented in the dressed-atom picture~\cite{CohenTannoudjiReynaud77}.

The double-well potentials~\cite{GatiOberthaler07} used for the single-particle experiments~\cite{KierigEtAl08} would
promise further exciting experimental results for BECs: a simple mean-field model of this system is known to be chaotic for
not too small
interactions~\nocite{UtermannEtAl94}\nocite{AbdullaevKraenkel00}\nocite{GhoseEtAl01}\nocite{LeeEtAl01}\cite{UtermannEtAl94,GhoseEtAl01,LeeEtAl01,AbdullaevKraenkel00,LasPhys08}.
For BECs in a double-well potential with periodic shaking, we demonstrated numerically the
emergence of ``Schr\"odinger-cat'' like states in phase-space 
if the parameters are chosen such that the mean-field system
displays a coexistence of regular and chaotic dynamics~\cite{WeissTeichmann08}.
In this paper,
more general entangled states are investigated. As shown by the authors of
Ref.~\cite{PezzeSmerzi2007}, the quantum Fisher information~\cite{BraunsteinEtAl96,BraunsteinCaves94} can be used to identify
multi-particle entanglement. This leads to a a greater variety of highly entangled states than
we investigated in Ref.~\cite{WeissTeichmann08}.

This manuscript is organised as follows: Section~\ref{sec:mod} introduces the models used to
describe the BEC in a double well both on the mean-field (Gross-Pitaevskii) level and on the
$N$-particle quantum level. Section~\ref{sec:multi} explains the way entangled states are
detected in the numerics. Before concluding the paper, Sec.~\ref{sec:results} shows the
numerical results for the emergence of entanglement.

\section{Model\label{sec:mod}}
Bose-Einstein condensates in double-well potentials are often described via a model
originally developed in nuclear physics~\cite{LipkinEtAl65}, a multi-particle Hamiltonian in two-mode
approximation~\cite{MilburnEtAl97}: 
\begin{eqnarray}
\label{eq:H}
\hat{H} &=& -\frac{\hbar\Omega}2\left(\hat{a}_1^{\phantom\dag}\hat{a}_2^{\dag}+\hat{a}_1^{\dag}\hat{a}_2^{\phantom\dag} \right) + \hbar\kappa\left(\hat{a}_1^{\dag}\hat{a}_1^{\dag}\hat{a}_1^{\phantom\dag}\hat{a}_1^{\phantom\dag}+\hat{a}_2^{\dag}\hat{a}_2^{\dag}\hat{a}_2^{\phantom\dag}\hat{a}_2^{\phantom\dag}\right)\nonumber\\
&+&\hbar\big(\mu_0+\mu_1\sin(\omega t)\big)\left(\hat{a}_2^{\dag}\hat{a}_2^{\phantom\dag}-\hat{a}_1^{\dag}\hat{a}_1^{\phantom\dag}\right)\;,
\end{eqnarray}
where the operator $\hat{a}^{(\dag)}_j$ creates (annihilates) a boson in well~$j$; $\hbar\mu_0$ is
the tilt between well~1 and well~2 and $\hbar\mu_1$ is the driving amplitude. The interaction
between a pair of particles in the same well is denoted by $2\hbar\kappa$.
Applications of such Hamiltonians include
multi-particle entanglement~\cite{MicheliEtAl03,MahmudEtAl03,TeichmannWeiss07,Creffield2007}, high precision
measurements, many-body quantum coherence~\cite{Lee06,LeeEtAl08} and spin
systems~\cite{DusuelVidal05}.

Within the approximation on which the Gross-Pitaevskii equation is based, a multi-particle
wave-function is characterised by two variables,  $\theta$ and $\phi$. The fraction of atoms
found in well~1 (well~2) is given by $\cos^2[\theta/2]$ ($\sin^2[\theta/2]$); the
phase-factor between both wells is given by $\exp(i\phi)$. All atoms occupy this single
particle wave-function. One can also find corresponding $N$-particle wave-functions 
known as ``atomic coherent states''~\cite{MandelWolf95}. In an expansion in the Fock-basis $|
n, N-n \rangle$ with $n$ atoms in well~$1$ and  $N-n$ atoms in  well~$2$ these wave-functions read:
\begin{eqnarray}
\label{eq:atomic}
\left|\theta,\phi\right>&=& \sum_{n=0}^N \genfrac{(}{)}{0pt}{}{N}{n}^{1/2}\cos^{n}(\theta/2)
                 \sin^{N-n}(\theta/2)
                 \nonumber\\
                 &\times& e^{i(N-n)\phi}| n, N-n \rangle\;.
\end{eqnarray}

The Gross-Pitaevskii dynamics can be mapped to that  of a nonrigid
pendulum~\cite{SmerziEtAl97}. Including the term  describing the  periodic shaking, the 
Hamilton function is given by ($z=\cos^2(\theta/2)-\sin^2(\theta/2)$):
\begin{eqnarray}
H_{\rm mf}& = &\frac{N\kappa}{\Omega}z^2-\sqrt{1-z^2}\cos(\phi)\nonumber\\
&-&2z\left(\frac{\mu_0}{\Omega}+
\frac{\mu_1}{\Omega}\sin\left({\textstyle\frac{\omega}{\Omega}}\tau\right)\right)\;,\quad \tau =t\Omega\;.
\label{eq:mean}
\end{eqnarray}
In our case, $z/2$ is 
the experimentally measurable~\cite{GatiOberthaler07} population imbalance which can
be used
to characterise the mean-field dynamics. The dynamics for this system are known to become
chaotic~\cite{GuckenheimerHolmes83}.

\section{Multi-particle-entanglement \& Quantum Fisher information\label{sec:multi}}

\begin{figure}
\vspace{-1cm}
\includegraphics[angle=-90,width=\linewidth]{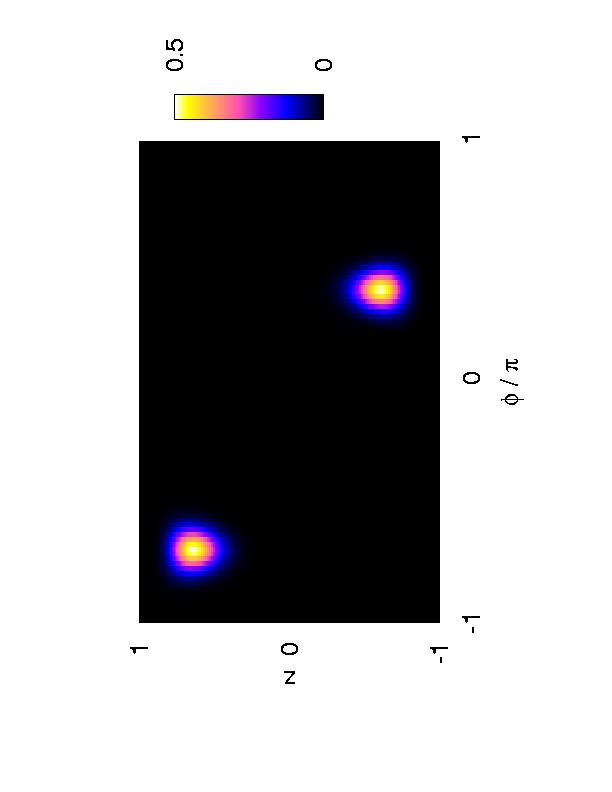}
\caption{\label{fig:cat}(Colour online) An ideal ``Schr\"odinger-cat'' like state which is the superposition of
  two atomic coherent states~(\ref{eq:atomic}) with hardly any overlap. 
The figure
shows the projection 
$|\langle \psi_{\rm cat}| \theta, \phi\rangle|^2$
 of the
``Schr\"odinger-cat'' like state $\psi_{\rm cat} = \frac{1}{\sqrt{2}} \left( |z=-0.6, \phi = 1.2 \rangle + |z = 0.65, \phi=-2.20 \rangle
\right)$ onto the atomic
coherent states $|\theta, \phi \rangle$ (e.q. \ref{eq:atomic}) in dependece of
$z$ and $\phi$.
While
  Ref.~\cite{WeissTeichmann08} concentrated on identifying such highly entangled state, the
  present manuscript uses a different approach to search highly entangled states: the sufficient condition~(\ref{eq:sufficient}) derived in
  Ref.~\cite{PezzeSmerzi2007} by using the quantum Fisher information~(\ref{eq:fisher})
  (Ref.~\cite{PezzeSmerzi2007} and references therein).}
\end{figure}

Multi-particle-entanglement~\cite{Yukalov2003,Vaucher08,DunninghamEtAl05,PoulsenEtAl05,CiracEtAl98,MicheliEtAl03,MahmudEtAl03,Dounas-frazerEtAl07}
is a hot topic of current research; to experimentally realise ``Schr\"odinger-cat'' like
superpositions of BECs is still a challenge of fundamental research. For the periodically driven
double-well potential, the relation between emergence of ``Schr\"odinger-cat'' like mesoscopic superpositions
in phase-space and mean-field chaos has been discovered in Ref.~\cite{WeissTeichmann08}. An ideal
example of such a mesoscopic superposition is shown
in Fig.~\ref{fig:cat}; the fidelity~\cite{fidelity} of some of the highly entangled states found numerically
was well above 50\%~\cite{WeissTeichmann08}. In this manuscript, we employ the fact that the
quantum Fisher information can be used to detect multi-particle entanglement~\cite{PezzeSmerzi2007}.
 
Before defining the quantum Fisher information, we note that the creation and annihilation
operators can be used to define
\begin{eqnarray}
\hat{J}_x =& &\frac12\left(\hat{a}^{\dag}_1\hat{a}_2^{\phantom{\dag}}+\hat{a}^{\dag}_2\hat{a}^{\phantom{\dag}}_1\right)\;,\\
\hat{J}_y =& -&\frac i2\left(\hat{a}_1^{\dag}\hat{a}_2^{\phantom{\dag}}-\hat{a}_2^{\dag}\hat{a}^{\phantom{\dag}}_1\right)\;,\\
\hat{J}_z =& &\frac 12\left(\hat{a}_1^{\dag}\hat{a}_1^{\phantom{\dag}}-\hat{a}_2^{\dag}\hat{a}_2^{\phantom{\dag}}\right)\;,
\end{eqnarray}
which satisfy angular momentum commutation rules. Except for a factor of $N$, the operator $\hat{J}_z$ is the operator
of the (experimentally measurable~\cite{GatiOberthaler07}) population imbalance.

While the quantum Fisher information~$F_{\rm Q}$ can be defined for statistical mixtures, it is
particularly simple for pure states~\cite{PezzeSmerzi2007}:
\begin{equation}
F_{\rm Q} = 4(\Delta \hat{J}_n)^2\;;
\end{equation}
where $\Delta \hat{J}_n$ are the mean-square fluctuations of $\hat{J}$ in direction $n$. In
the following we choose the $z$-direction, thus
\begin{equation}
\label{eq:fisher}
F_{\rm Q} = 4(\Delta \hat{J}_z)^2\;.
\end{equation}
For $N$ particles, a sufficient condition for multi-particle entanglement is given
by~\cite{PezzeSmerzi2007}:
\begin{eqnarray}
\label{eq:sufficient}
F_{\rm ent}&>&1\;,\\
F_{\rm ent}&\equiv&\frac{F_{\rm Q}}N\;.
\label{eq:flag}
\end{eqnarray}
For the ideal ``Schr\"odinger-cat'' state in real space,
\begin{equation}
|\psi_{\rm NOON}\rangle=\frac1{\sqrt{2}}\left(|N,0\rangle+|0,N\rangle\right)\;,
\end{equation}
this entanglement flag reaches a value of $N$. Thus, while Eq.~(\ref{eq:sufficient}) already
indicates multi-particle entanglement, values of 
\begin{equation}
\label{eq:highly}
F_{\rm ent}\gg 1
\end{equation}
 demonstrate highly entangled
states.

\section{Results\label{sec:results}}
\begin{figure}[th]
\includegraphics[angle=-90,width=\linewidth]{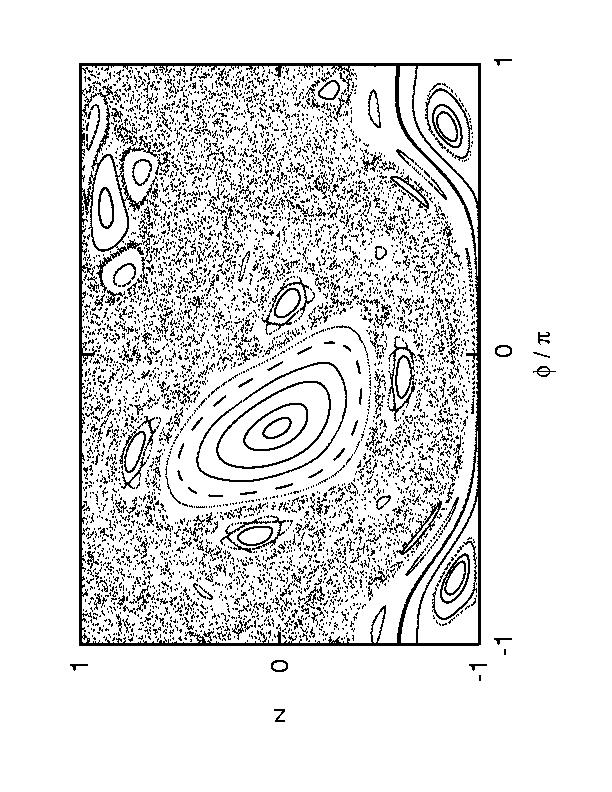}
\caption{\label{fig:sosa}Poincar{\'e} surface of section for the mean-field system. Closed
  loops characterise stable orbits whereas chaos is represented by irregular dots. The
  parameters are chosen such that they correspond to a one-photon resonance:  a tilt of $2\mu_0/\Omega=3.0$, a driving frequency of~$\omega=3\Omega$, an interaction of $N\kappa/\Omega=0.8$ and a driving amplitude of $2\mu_1/\Omega=0.9$ (cf.\ Ref.~\cite{EckardtEtAl05}).}
\end{figure}

In order to characterise whether or not the mean-field dynamics are chaotic, Poincar\'e
surface of sections are particularly suited: for a set of initial conditions, the mean-field
dynamics of the periodically driven system characterised by the angular frequency $\omega$ is
plotted at integer multiples of $2\pi/\omega$. Figure~\ref{fig:sosa} shows a phenomenon which
is very characteristic for the non-rigid pendulum on which the BEC in a double well was
mapped within mean-field: the coexistence between chaotic and  regular regions.

\begin{figure}
\vspace*{-1cm}
\includegraphics[width=\linewidth]{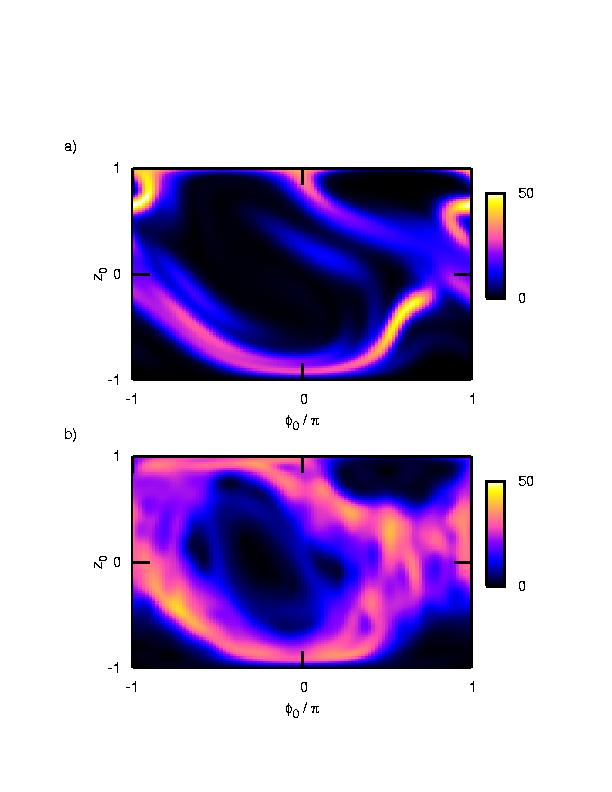}
\caption{\label{fig:entfig1a}(Colour online) The entanglement flag~(\ref{eq:flag}) in a two-dimensional
  projection as a function of the initial condition for $N=100$ particles. All other
  parameter as in Fig.~\ref{fig:sosa}. The $N$-particle wave-function which corresponds to
  the mean-field initial conditions ($z_0$,$\phi_0$) can be found in
  Eq.~(\ref{eq:atomic}). In the upper panel, the (experimentally measurable)  entanglement
  flag~(\ref{eq:flag}) is shown after a time of $t\Omega=10$, in the lower panel the time is
  $t\Omega=100$. Highly entangled states (cf.\ Eq.~(\ref{eq:highly})) occur in the entire
  chaotic regime.
  }
\end{figure}

 While each
initial condition will, in general, lead to many dots in plots like Fig.~\ref{fig:sosa}, we
proceed in the spirit of Ref.~\cite{WeissTeichmann08} to characterise the $N$-particle
dynamics: In Fig.~\ref{fig:entfig1a} each initial condition leads to only one point: the
entanglement flag~(\ref{eq:flag}) for this initial condition after a fixed time~$t\Omega$. Highly entangled states
(Eq.~(\ref{eq:highly})) can be found on the boundary of stable regions; already for short
times (Fig.~\ref{fig:entfig1a}~a) such highly entangled states can occur. For larger times
(Fig.~\ref{fig:entfig1a}~b) many features of the Poincar\'e surface of section in
Fig.~\ref{fig:sosa} are visible in the entanglement generation which essentially spreads over
the entire chaotic part of the initial conditions.


\begin{figure}
\includegraphics[angle=-90,width=\linewidth]{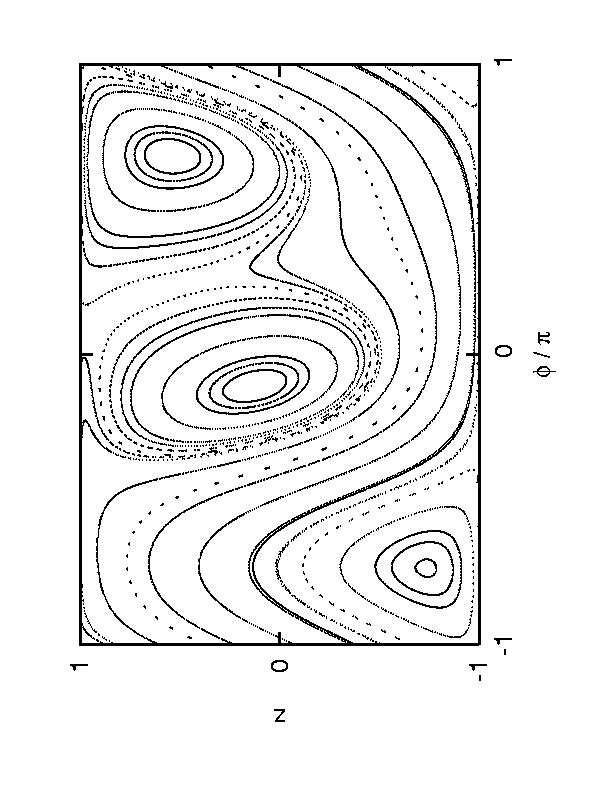}
\caption{\label{fig:sosb}Poincar{\'e} surface of section for the mean-field system (cf.\
  Fig.~\ref{fig:sosa}).  The
  parameters are chosen such that they correspond to a $3/2$-photon resonance~\cite{EckardtEtAl05}  with
  $N\kappa/\Omega=0.1$,  $2\mu_0/\Omega=3.0$, $\omega/\Omega=2.08$ and~$2\mu_1/\Omega=1.8$.}
\end{figure}

\begin{figure}
\vspace*{-2cm}
\includegraphics[width=\linewidth]{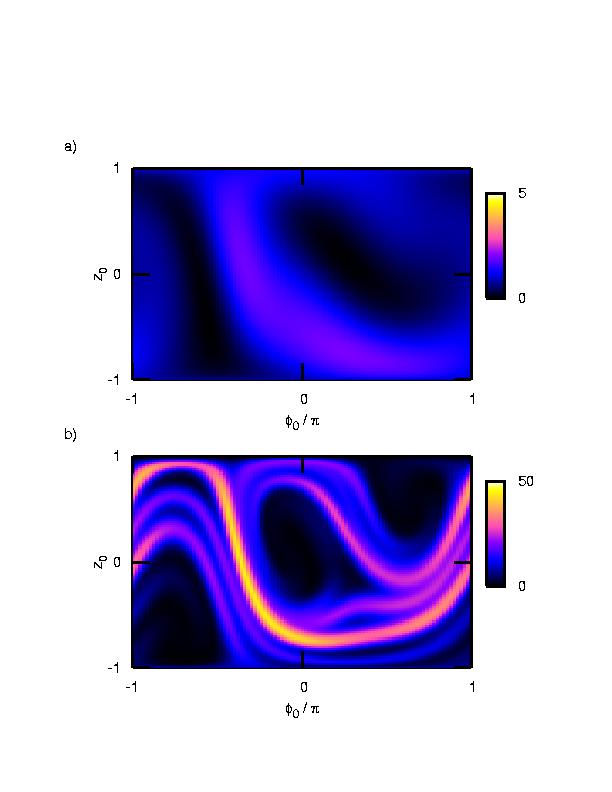}
\caption{\label{fig:entfig1b}(Colour online) The entanglement flag~(\ref{eq:flag}) in a two-dimensional
  projection as a function of the initial condition (cf.\ Fig.~\ref{fig:entfig1a}) for $N=100$ particles; all other
  parameter as in Fig.~\ref{fig:sosb}. For the regular regime given by the parameters of
  Fig.~\ref{fig:sosb}, after short time-scales of $t\Omega=10$ (upper panel)
  entanglement-production is very weak (note that the brightness-entanglement-coding differs by
  a factor of 10 from Fig.~\ref{fig:entfig1b} and the longer time-scales $t\Omega=100$
  depicted in the lower panel). For larger time-scales, entanglement production mainly occurs
on the boundary of stable regions (lower panel).}
\end{figure}

For parameters which display no chaotic parts in the Poincar\'e surface of section
(Fig.~\ref{fig:sosb}), on short time-scales hardly any entanglement emerges
(Fig.~\ref{fig:entfig1b} a). For larger time-scales, entanglement generation occurs on the
boundaries of stable  regions (Fig.~\ref{fig:entfig1b} b).

\section{Conclusion}

In Ref.~\cite{WeissTeichmann08} we discovered the relation between mesoscopic ``Schr\"odinger-cat''
like superpositions in phase space and mean-field entanglement. In the present paper, we
demonstrated that also for more general entangled states, multi-particle entanglement can be
a quantum signature of chaos. For regular systems, the general entangled states also
occur. However, it is restricted to the boundaries of stable regions and only occurs on
longer time-scales. While the focus in
Ref.~\cite{WeissTeichmann08} was on finding particularly highly entangled states for each
initial condition, in the present manuscript the entanglement production was shown for each
initial condition at the same point in time, both for short and longer times between the
onset of the computer experiment and the read-out. 

As an entanglement-flag, we apply the quantum Fisher information for pure states.
In this paper we use it in a way which is
particularly easy to measure experimentally. However, using Eq.~(\ref{eq:flag}) assumes a pure state and
would thus only be valid in an ideal system without decoherence. For
more realistic situations, 
experimental signatures as in Refs.~\cite{PiazzaEtAl2008,WeissCastin08} should be
investigated in the future.

\acknowledgments

We would like to thank A.~Smerzi for useful discussions and M.~Holthaus for his
continuous support. N.T.\ acknowledges funding by the Studienstiftung des deutschen Volkes.


\begin{thebibliography}{50}
\bibitem[{Las()}]{LasPhys08}
\bibinfo{note}{To investigate chaos, systems like the delta-kicked
  rotor~\cite{Moore95,dArcy01,CreffieldEtAl06,StrzysEtAl08} have also been
  used. Research topics range from quantum signatures of chaos~\cite{Haake92}
  and quasi-stationary distributions~\cite{BreuerEtAl00} over
  entanglement~\cite{Garcia-mata07,GhoseSanders04} to
  decoherence~\cite{Braun01}.}
\bibitem[{fid()}]{fidelity}
\bibinfo{note}{The fidelity of a highly entangled state is defined as the
  probability to be in an ideal target state. For Schr\"odinger-cat like
  states, only fidelities above 50\% are physically significant.}
\end{thebibliography}
\end{document}